\title{\hspace{0.1 cm} The Reactive Volatility Model  }

\author{Sebastien Valeyre$^1$ \and Denis Grebenkov$^2$ \and Sofiane Aboura$^3$  \and Qian Liu$^4$ \\
\\
$^{1,4}$ John Locke Investments \\ 38 Avenue Franklin Roosevelt, 77210 Fontainebleau-Avon, France.
 \\ Emails: sebastien.valeyre@jl-investments.com - qian.liu@jl-investments.com \\
\\
$^2$ Laboratory of Condensed Matter Physics, CNRS -- Ecole Polytechnique \\ F-91128 Palaiseau, France. \\ Email: denis.grebenkov@polytechnique.edu \\
\\
$^3$ Universit\'e de Paris-Dauphine, DRM-Finance \\ Place du Maréchal de Lattre de Tassigny, 75775 Paris cedex 16, France. \\ Email: sofiane.aboura@dauphine.fr
}

\date{\today}

\documentclass[11pt]{article} 
\usepackage{amsmath}
\usepackage{amsfonts}
\usepackage[utf8]{inputenc}
\usepackage{amssymb}
\usepackage{graphicx}
\usepackage[english]{babel}
\usepackage[T1]{fontenc} 
\usepackage{ragged2e}
\usepackage[tmargin=1.5in, bmargin=1.5in, lmargin=1.0in, rmargin=1.0in]{geometry}
\usepackage{multirow}

\usepackage{natbib}

\usepackage{color}

\def\P{{\mathbb P}}

\begin{document}
\maketitle

\begin{abstract}
\justifying
We present a new volatility model, simple to implement, that 
includes a leverage effect whose return-volatility correlation
function fits to empirical observations.  This model is able to
capture both the ``retarded effect'' induced by the specific risk, and
the ``panic effect'', which occurs whenever systematic risk becomes
the dominant factor.  Consequently, in contrast to a GARCH model and a
standard volatility estimate from the squared returns, this new model
is as reactive as the implied volatility: the model adjusts itself in
an instantaneous way to each variation of the single stock price or
the stock index price and the adjustment is highly correlated to
implied volatility changes.  We also test the reactivity of our model
using extreme events taken from the 470 most liquid European stocks
over the last decade.  We show that the reactive volatility model is
more robust to extreme events, and it allows for the identification of
precursors and replicas of extreme events.
\\ Key Words: Volatility, tail event, risk management, asset pricing
\\ JEL Classification: C5, G01, G1, G32
\\
\end{abstract}

\section{Introduction}

Stylized facts from the financial markets include heavy tails, extreme
correlation and leverage effect
\citep{Bouchaud00,Bouchaud01b,Cont01}.  Various studies suggest
that the power law nature of financial returns, $\P\{ |r| > x\}
\propto x^{-\alpha}$ (at large $x$) with a quasi-universal exponent
$\alpha$ close to $3$, explains most extreme events, including crashes
\citep{Gabaix03}.  All these extreme events in the time series affect
the estimation of the tail dependence measure \citep{Davis09}.  In
particular, when the market is facing extreme negative returns, the
level of correlation is significantly higher than during extreme
positive returns \citep{Longin01,Ang02}.  This pattern can be
explained by the asymmetry of volatility that is partly due to the
so-called ``leverage effect''.  The leverage effect is characterized
by a surge in volatility and a subsequent drop in the stock price
\citep{Black76,Christie82,Campbell92,Bekaert00}.  

The leverage effect puzzle is well documented in recent
literature \citep{Perello03,Bollerslev06,Qiu06,Cevdet07,Ait13}.
Various approaches have been suggested to include the leverage
effect.

{\bf Continuous-time models}.  The Constant Elasticity of Variance
model (CEV) developed by \citet{Cox75} is the first model that
explicitly describes the relation between volatility and price.  The
elasticity stock price exponent allows the instantaneous variance of
the percentage price change to be a direct inverse function of the
stock price.  \citet{Harvey96} developed a stochastic volatility model
with leverage effect to be estimated by a quasi-maximum likelihood
procedure.  \citet{Hagan00} proposed a generalization of the CEV
dynamics known as the SABR model.  \citet{Carr09} proposed to model
the equity-to-asset ratio by a simple CEV process in which large
discontinuous shocks are modeled as a jump process to capture a
self-exciting behavior.  \citet{Veraart09} modeled the leverage effect
as a stochastic process by introducing an additional source of
randomness into the stochastic volatility model.  More recently,
\citet{Filimonov11} developed a model describing critical events in
the self-organized systems such as financial markets.  It relies on a
self-excited multifractal process with an explicit dependence of the
dynamics of the process on both external events and internal memory.
It allows for leverage effect as an intrinsic property.

{\bf Discrete-time models}.  In the ARCH literature, the
conditional variance is specified to be a function of the size and the
sign of returns.  One can mention the Exponential-GARCH
\citep{Nelson91}, the GJR-GARCH \citep{Glosten93}, the Asymmetric
Power-ARCH \citep{Ding93}, Nonlinear-GARCH \citep{Engle93} and the
Threshold-GARCH \citep{Zakoian94}.

To our knowledge, both the continuous-time and discrete-time
approaches remain focused mainly on theoretical issues.
\citet{Bouchaud01} are the first to study in detail the dynamics of
the leverage effect adapted to stocks and stock indices.  In
particular, they introduced and measured the ``return-volatility
correlation function'' for stocks and indices (Fig. \ref{fig:MC}).
They derived two different dynamics for the systematic and specific
risk and found
a moderate (strong) correlation with a decay period of 50 (10) days
for individual stocks (indices).  The correlations for indices are
therefore stronger than that for stocks, despite the fact that a stock
index is merely a portfolio of stocks. They argue that the leverage
effect for stocks stems from a simple retarded effect, in which price
variations are calibrated not on the instantaneous value of the price
but on an exponential moving average of the price.  Their retarded
volatility model adequately represents the leverage effect for
individual stocks, which are mainly characterized by idiosyncratic (or
specific) risks.  However, they recognize that it is no longer the
case for stock indices, which are only characterized by systematic
risk, whose leverage is driven by another phenomenon, the so-called
``panic effect''.  This panic effect is partly explained by an
increase of correlation between single stocks \citep{Reigneron11}.
In fact, a fast decline of the market leads to the panic of
investors and to an increase of the systematic risk of each single
stock which may become the dominant factor when the panic is extreme.
As emphasized by Allez and Bouchaud (2011), ``during large swings of
the index, the market exposure of stocks becomes the dominant
factor''.

Relying on the retarded volatility model by \citet{Bouchaud01}, we
develop a new volatility model, simple to implement, that (i)
conserves the retarded effect property and (ii) adequately extends the
leverage effect from the specific single stock case to the systematic
case of stock indices, taking into account the panic effect.  This
model is well suited not only for stock indices but also for single
stocks, particularly in times of distress during which stocks are
mainly affected by the systematic risk.  This advantage enables
the model to be as reactive as the implied volatilities, i.e. there is
no delay between the implied volatility changes and our model: they
remain highly correlated.  This important property makes it different
from most previously published theoretical models.

The reactivity of the model is first tested against the European
volatility index V2X and is also compared with two classical models
selected as benchmarks: the GARCH model and the standard volatility
estimate based on the square root of the exponential moving average of
squared returns.  The comparison to the volatility index is chosen
because volatility is not directly observable and market participants
prefer using implied volatility indices as market volatility proxies.
Thus, how well the model captures the dynamics of such a proxy may be
an adequate gauge of quality.

The robustness of the model is then tested using extreme events.  An
empirical study is performed on the 470 most liquid European stocks
over the last decade.  We investigate extreme systematic and specific
risks, which could be responsible for massive losses and are therefore
important for investors.  Our results suggest that the market shocks
are better assimilated into the reactive volatility model.

The article is organized as follows.  Section \ref{sec:RVM} describes
the reactive volatility model.  Section \ref{sec:empirical} analyzes
the empirical robustness of the model near extreme events.  Section
\ref{sec:conclusion} summarizes the main findings and provides some
concluding remarks.

\section{The reactive volatility model}
\label{sec:RVM}

The reactive volatility model takes into account two different dynamic
features of the leverage effect: it models the panic effect related to
the systematic risk and combines it with the retarded volatility model
of \citet{Bouchaud01} that accurately describes the slow dynamics of
the leverage effect for the specific risk.  We focus first on the
stock index case since it allows us to introduce the model of the
panic effect that governs the systematic risk.  Second we treat the
single stock case that combines both the specific and systematic
risks.  We show how the retarded volatility model is combined with the
panic effect in order to model at the same time the leverage effect of
the specific and systematic risks.

\subsection{Model of the panic effect describing the systematic risk: the case of stock index}
\label{sec:stock_index}

Let $I(t)$ be a stock index at equally spaced, discrete times $t$.  It
is well known that arithmetic returns, $\Delta I(t)/I(t)$, are
heteroscedastic, partly due to price-volatility correlations.  The
goal is to define a convenient estimator, a level $L(t)$, of the stock
index $I(t)$ such that the renormalized arithmetic returns, $\Delta
I(t)/L(t)$, become more homoscedastic.
%
%
\begin{figure}
\begin{center}
\includegraphics[width=100mm]{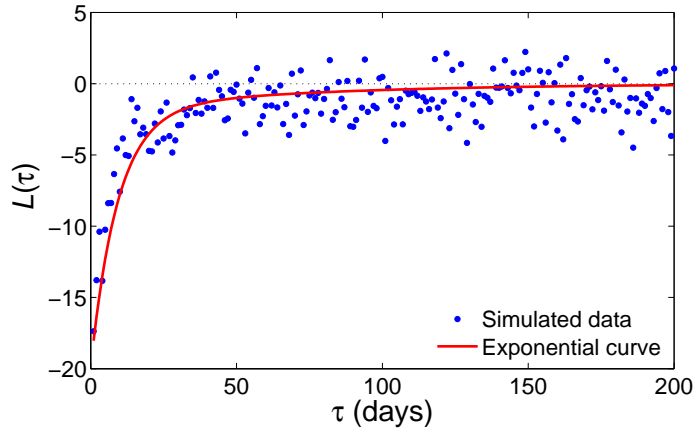}
\end{center}
\caption{
The return-volatility correlation function (leverage effect),
${\mathcal L}(\tau) \equiv \langle [\delta x(t+\tau)]^2 \delta x(t)
\rangle/\langle [\delta x(t)]^2\rangle^2$, with $\delta x(t) = \Delta
I(t)/I(t)$, for simulated data (symbols), and the exponential curve
${\mathcal L}(\tau) \approx - A_I \exp(-\tau/T_I)$, with $A_I = 18$
and $T_I = 9.3$ (days), as introduced and calibrated by
\citet{Bouchaud01}.  The simulated data are obtained from a Monte Carlo
generated price series of 10,000 points, with $I(t+1) = I(t) (1 +
\sigma_I(t) \varepsilon_t)$, in which i.i.d. $\varepsilon_t\sim
{\mathcal N}(0,1)$, and $\sigma_I(t)$ is given by
Eq. (\ref{eq:sigmaIt1}) with a constant $\tilde{\sigma}_I(t) = 0.01$.
This figure supports the claim that the reactive volatility model
captures the leverage effect, using the correct parameters $A_I$ and
$T_I$ calibrated from the empirical stock indices.}
\label{fig:MC}
\end{figure}

Let us introduce two stock index levels as exponential moving averages
(EMAs) with two characteristic time scales: a slow level $L_s(t)$ and
a fast level $L_f(t)$.  These EMAs can be computed using standard
linear relations:
\begin{eqnarray}
L_s(t+1) &=& (1-\lambda_s) L_s(t) + \lambda_s I(t+1) \\
L_f(t+1) &=& (1-\lambda_f) L_f(t) + \lambda_f I(t+1)
\end{eqnarray}
where $\lambda_s$ and $\lambda_f$ are the weighting parameters of the
EMAs.  The appropriate values of $\lambda_s = 0.0241$ and $\lambda_f =
0.1484$ are extracted from the measurement of the
return-volatility correlation function in \citet{Bouchaud01}, see the
caption of Fig. \ref{fig:MC} for details.  The slow parameter
corresponds to the relaxation time of the retarded effect for the
specific risk whereas the fast one corresponds to the relaxation time
of the panic effect for the systematic risk.  These two relaxing times
are found rather universal as they are stable in time and remain close
to each other for different mature stock markets.

For practical purposes, a filter is introduced to make the estimator
more robust against outliers or extreme instantaneous variations of
the stock index.  We set:
\begin{equation}
\hat{L}_s(t+1) = I(t+1) \left(1 + F_\phi\left(\frac{L_s(t+1) - I(t+1)}{I(t+1)}\right)\right)
\end{equation}
where a filter function $F_\phi(z)$ is proportional to $z$ for small
$z$ and saturated to a constant for large $|z|$.  We expect that
the leverage effect is linear up to a certain point.  We choose
$\displaystyle F_\phi(z) = \frac{\tanh(\phi z)}{\phi}$,
where $\phi$ is a parameter that determines the region of linearity of
the filter, i.e., smaller $\phi$ correspond to wider linearity
regions.%
\footnote{We set $\phi = 1/0.3 \approx 3.3$, which corresponds to a
maximum stock index daily variation of $\pm 30\%$, or a maximum
drawdown in the order of $30\%$ over $1/\lambda_s \approx 40$ days.
For example, during the worst American stock market crash on 19
October 1987, the S\&P 500 declined by $22\%$ while the VIX climbed up
to $150\%$.}
At $\phi = 0$, there is no filter, $F_0(z) = z$ and $\hat{L}_s(t+1) =
L_s(t+1)$.  This filter has only a very minor impact on the
results and is useful in practice to level off a couple of very
extreme events.  One could use any other S-shaped function which is
expected to give similar results.
Finally, the main equation for the stock index level $L(t)$, in which
the fast level is modulated by the filtered slow level, is defined as:
\begin{equation}
\label{eq:L}
L(t+1) = \hat{L}_s(t+1) \left(1 + F_\phi \left(\left(\frac{L_f(t+1)}{I(t+1)}\right)^\ell - 1\right)\right)
\end{equation}
where $\ell$ is the leverage parameter that describes the amplitude of
the leverage effect between stock returns and volatility.  

The leverage parameter $\ell$ is set to $8$ to reproduce the double
exponential fit of the return-volatility correlation function in
the stock index case (see the caption of Fig. \ref{fig:MC} for
details).  For instance, the value of $\ell = 8$ means that if the
index varies by $1\%$, the volatility is expected to vary by
$-\ell\cdot 1\% = -8\%$.  That can be seen by using a Taylor
expansion: if there is no filter ($\phi = 0$) and $L_f(t+1)$ is close
to $I(t+1)$, a Taylor expansion of Eq. (\ref{eq:L}) yields a simpler
form:
\begin{equation}
\label{eq:L_Taylor}
L(t+1) \simeq \hat{L}_s(t+1) \left(1 + \ell ~ \frac{L_f(t+1) - I(t+1)}{L_f(t+1)}\right)
\end{equation}
%
%
\begin{figure}
\begin{center}
\includegraphics[width=100mm]{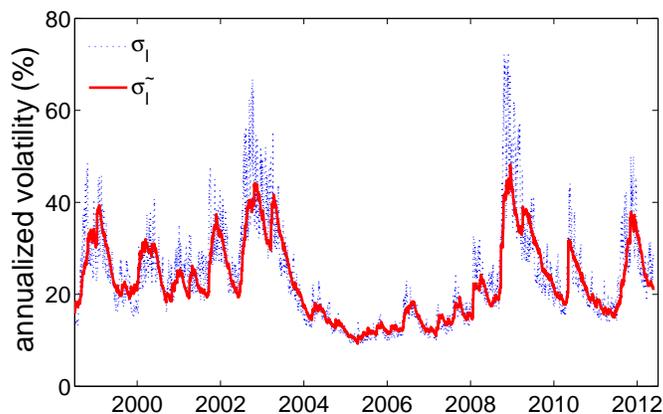}
\end{center}
\caption{
The measures of index volatility, $\sigma_I(t)$, from the reactive
volatility model (dashed blue line) and the renormalized volatility
$\tilde{\sigma}_I(t)$ (solid red line), for the period from 22/06/1998
to 18/05/2012.  The renormalized volatility is less volatile in the
short term but is able to capture all the long-term moves.  }
\label{fig:volatility}
\end{figure}

The Taylor expansion shows that the panic effect could be
modeled in a simple way with two exponential moving averages whereas
the retarded volatility model needs only one.  The level $L(t)$ is the
slow moving average of the retarded volatility model which is
modulated by the relative excess of the systematic risk $I(t+1)$ as
compared to its fast moving average $L_f(t+1)$.  This modulation is
amplified by the $\ell$ parameter which describes the ``capacity'' of
the market to panic: $\ell =0$ corresponds to a market that never
panics.  It seems that the $\ell$ parameter is also universal as it is
stable in time and is approximately the same for different mature
stock markets.

The introduction of the leverage effect into Eq. (\ref{eq:L}) allows
one to use the ``corrected'' level $L(t)$ instead of $I(t)$, which
leads to more homoscedastic stock index returns $\Delta I(t)/L(t)$
than $\Delta I(t)/I(t)$, as tested below.

An estimator of the renormalized variance $\tilde{\sigma}_I^2$ is
obtained through an EMA based on $L(t)$:
\begin{equation}
\label{eq:sigmaIt0}
\tilde{\sigma}_I^2(t+1) = (1 - \lambda_\sigma) \tilde{\sigma}_I^2(t) + \lambda_\sigma \left(\frac{\Delta I(t+1)}{L(t+1)}\right)^2
\end{equation}
where $\lambda_\sigma$ is a weighting parameter.  
$\lambda_\sigma$ has to be chosen as a compromise between the
estimation accuracy of the standard deviation of the renormalized
returns and the reactivity of that estimation.  Indeed the
renormalized returns are rather homoscedastic by construction in the
short time scale only (in fact, the renormalization based on the
leverage effect with short relaxation times ($\lambda_s$, $\lambda_f$)
cannot account for long time scale changes of volatility related to
economic cycles).  Since economic uncertainty does not change
significantly in a period of 2 months (40 trading days), we set
$\lambda_\sigma$ to $1/40 = 0.025$.  The related sample length leads
to a noise of 4\% in the annualized volatility which is in the order
of 20\%.

Figure \ref{fig:volatility} shows that $\tilde{\sigma}_{I}^{2}$ yields
a stable output in the short time scale since the variation of the
index $\Delta I(t+1)$ normalized by the index level $L(t+1)$ is rather
homoscedastic by construction.  The reactive volatility model is then
obtained through this stable output but using a reactive
renormalization factor. The reactivity of the model therefore comes
from the renormalization factor $L(t+1)/I(t+1)$ that will adjust to
every price move in an instantaneous way.  The reactive volatility
model $\sigma_I(t)$ for the systematic risk which is governed by the
panic effect (the stock index case) is defined as:
\begin{equation}
\label{eq:sigmaIt1}
\sigma_{I}(t+1) =\tilde{\sigma}_{I}(t+1) \frac{L(t+1)}{I(t+1)}
\end{equation}
Using the Taylor expansion (\ref{eq:L_Taylor}), we show that the
normalization factor depends mainly on the slow and fast EMAs of the
stock index price.  The reactive volatility is therefore obtained by
the following approximation:
\begin{equation}
\sigma_I(t+1) \approx \tilde{\sigma}_I(t+1) \frac{L_s(t+1)}{I(t+1)} \left(1 + \ell \frac{L_f(t+1)-I(t+1)}{L_f(t+1)}\right)
\end{equation}

\subsection{Volatility term structure and the empirical test against the V2X index}

We test the reactivity of Eq. (\ref{eq:sigmaIt1}) against the V2X
implied volatility.  Because the V2X index represents the implied
volatilities of the Eurostoxx 50 index with a maturity $T$ of one
month, one also needs to consider the volatility term structure of
Eq. (\ref{eq:sigmaIt1}) which is defined as:
\begin{equation}
\sigma_T^2(t) = \frac{1}{T} \langle \left(\int_{t}^{T+t} \sigma_{I}^2(t') dt' \right)\rangle
\end{equation}
where $\langle \ldots \rangle$ is the expectation (average) over
all possible trajectories of $\sigma_I(t')$ between $t$ and $t+T$.

\begin{figure}
\begin{center}
\includegraphics[width=120mm]{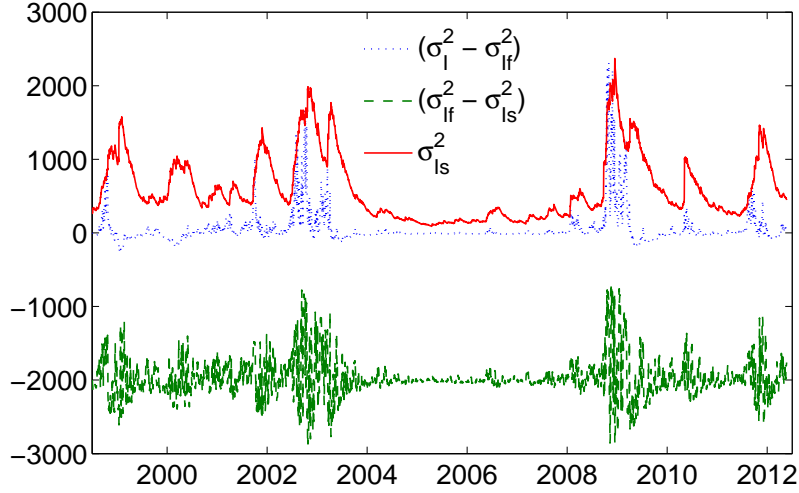}
\end{center}
\caption{
Empirical behavior of three ``components'' in Eq. (\ref{eq:sigmaI}):
$(\sigma_{I}^2(t) - \sigma_{I f}^2(t))$ (dotted blue line),
$(\sigma_{I f}^2(t) - \sigma_{I s}^2(t))$ (dashed green line, which is
shifted by $-2000$ along the vertical axis), and ${\sigma}_{I s}^2(t)$
(solid red line).  The first two components can be seen as
mean-reverting processes with different relaxation rates, while the
last varies much slower than the others. }
\label{fig:sigmaI}
\end{figure}

In order to approximate $\sigma_T^2(t)$, one could have implemented
a numerical solution based on Monte Carlo simulation (as described in
the caption of Fig. \ref{fig:MC}).  An alternative is to introduce an
empirical recipe based on a two-factor model with the following two
``long-term'' volatilities, $\sigma_{I s}(t)$ (the slow factor) and
$\sigma_{I f}(t)$ (the fast factor).  In this approximation, we assume
the instantaneous volatility $\sigma_I(t)$ to mean-revert toward the
fast ``long-term'' volatility $\sigma_{I f}(t)$, which in turn
mean-reverts toward the slow ``long-term'' volatility $\sigma_{I
s}(t)$.  The fast and slow ``long-term'' volatilities are defined as:
\begin{equation}
\sigma_{I s}(t) = {\sigma}_{I}(t) \frac{I(t)}{L_{s}(t)} \qquad   \sigma_{I f}(t) = {\sigma}_{I}(t) \frac{I(t)}{L_f(t)}
\end{equation}
We split the squared volatility $\sigma_I^2(t)$ into three
``components'':
\begin{equation}
\label{eq:sigmaI}
\sigma_{I}^2(t) = ({\sigma}_{I}^2(t) - {\sigma}_{I f}^2(t)) +  ({\sigma}_{I f}^2(t) - {\sigma}_{I s}^2(t)) + {\sigma}_{I s}^2(t)
\end{equation}
The empirical observation from Fig. \ref{fig:sigmaI} confirms, without
pretending to any rigor, that $(\sigma_{I}^2(t) - \sigma_{I f}^2(t))$
and $(\sigma_{I f}^2(t) - \sigma_{I s}^2(t))$ can be seen as two
mean-reverting processes with two relaxation rates $\lambda_{s}$ and
$\lambda_{f}$, while ${\sigma}_{I s}^2(t)$ varies much slower than the
other processes.  We therefore assume that $(\sigma_{I}^2(t) -
\sigma_{I f}^2(t))$ and $(\sigma_{I f}^2(t) - \sigma_{I s}^2(t))$ can
be approximated by Ornstein-Uhlenbeck processes with two relaxation
rates $\lambda_{s}$ and $\lambda_{f}$, while ${\sigma}_{I s}^2(t)$
follows Brownian motion.  These last assumptions (current in the
literature, e.g., \citet{Hull87,Heston93}) enable the reactive
volatility estimator with the term structure, $\sigma_T$, to be
estimated as:
\begin{equation}
\label{eq:sigmaT}
\sigma_T^2(t) \approx (\sigma_{I}^2(t) - \sigma_{I f}^2(t) ) \frac{(1-e^{-\lambda_{f}T})}{\lambda_{f}T} +
(\sigma_{I f}^2(t) - {\sigma}_{I s}^2(t)) \frac{(1-e^{-\lambda_{s}T})}{\lambda_{s}T}
+ \sigma_{I s}^2(t)
\end{equation}
This relation can be seen alternatively as an empirical definition of
$\sigma_T(t)$ or as a practical recipe for its computation.

\begin{figure}%
\begin{center}
\includegraphics[width=80mm]{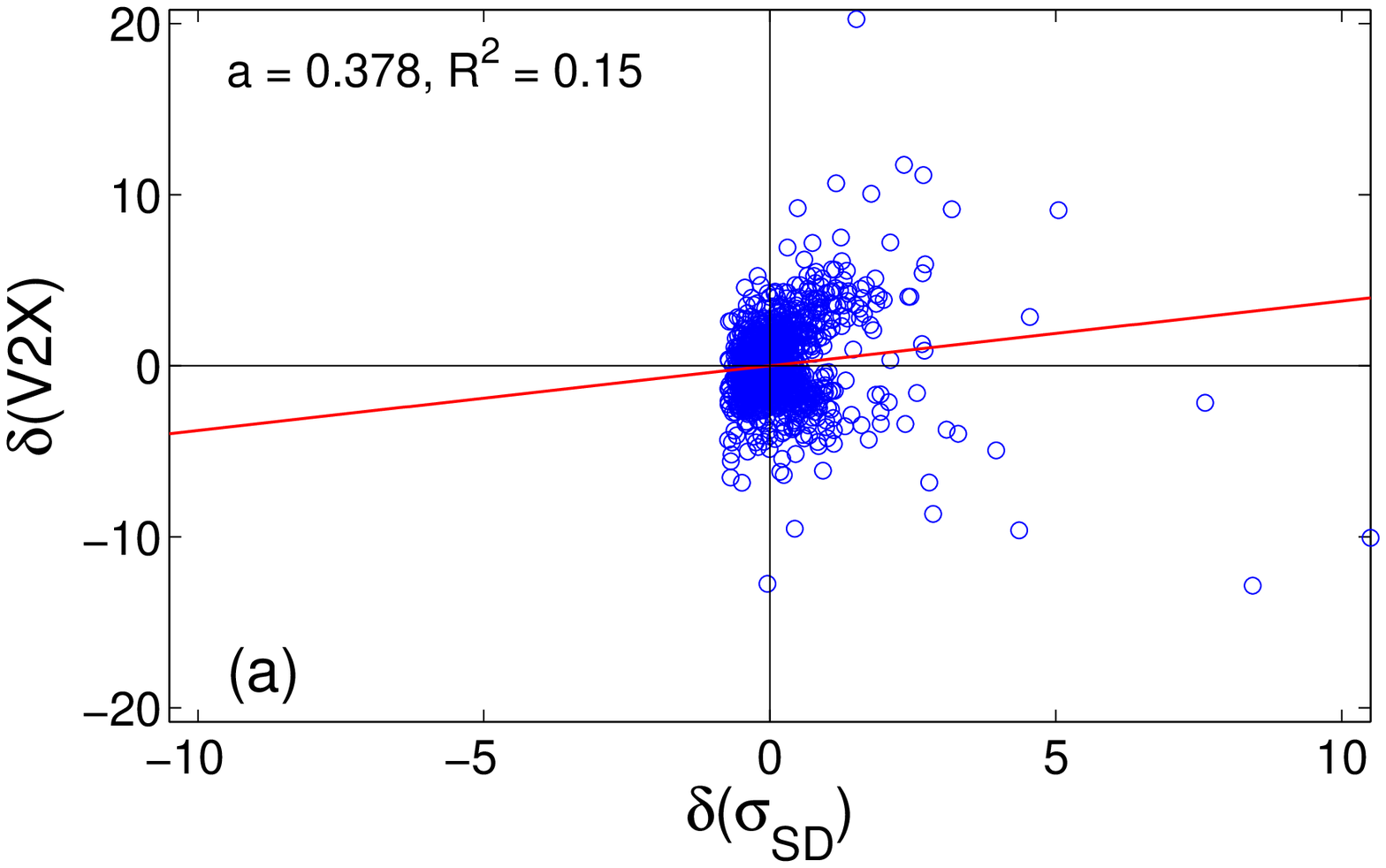}
\includegraphics[width=80mm]{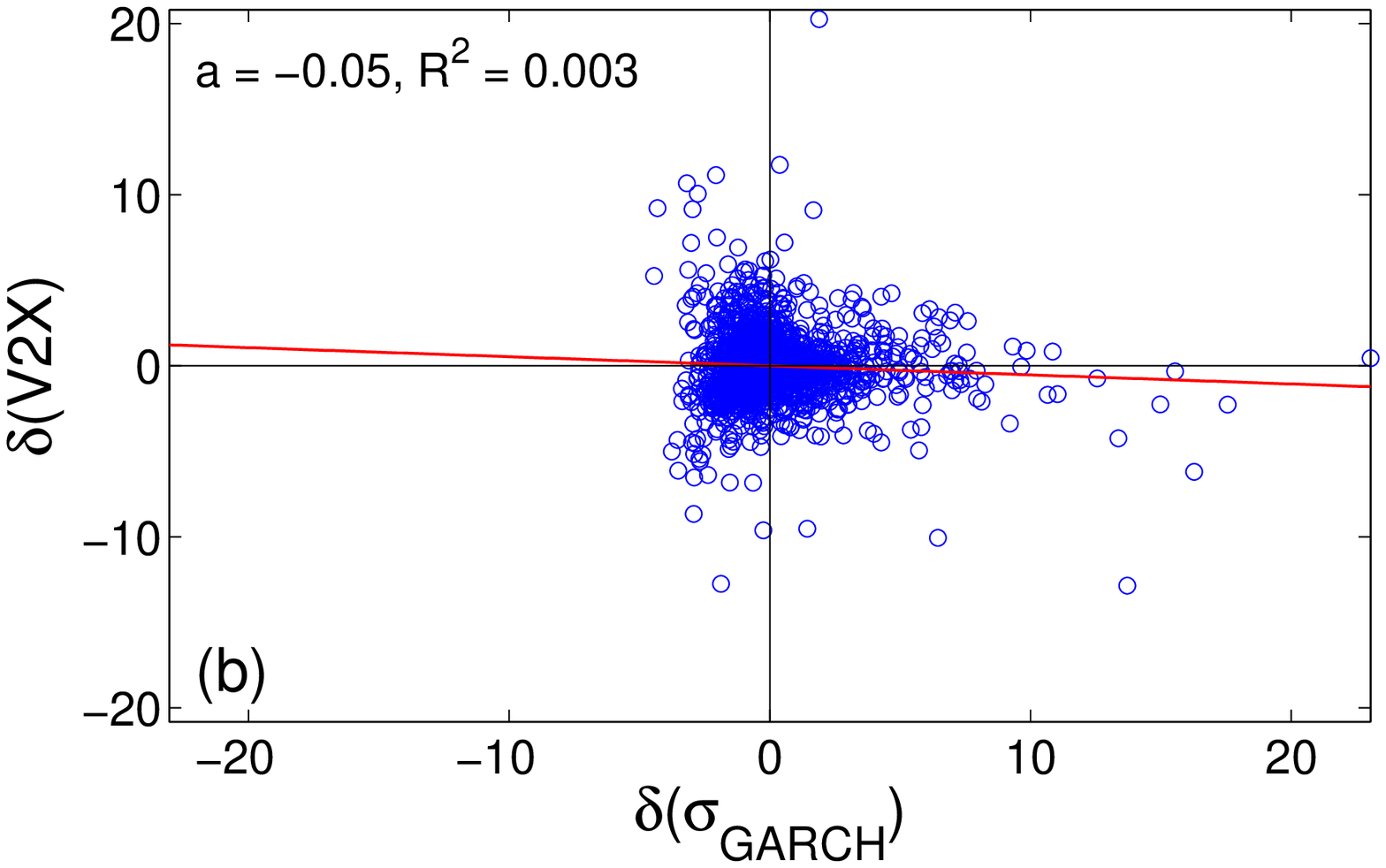}
\includegraphics[width=80mm]{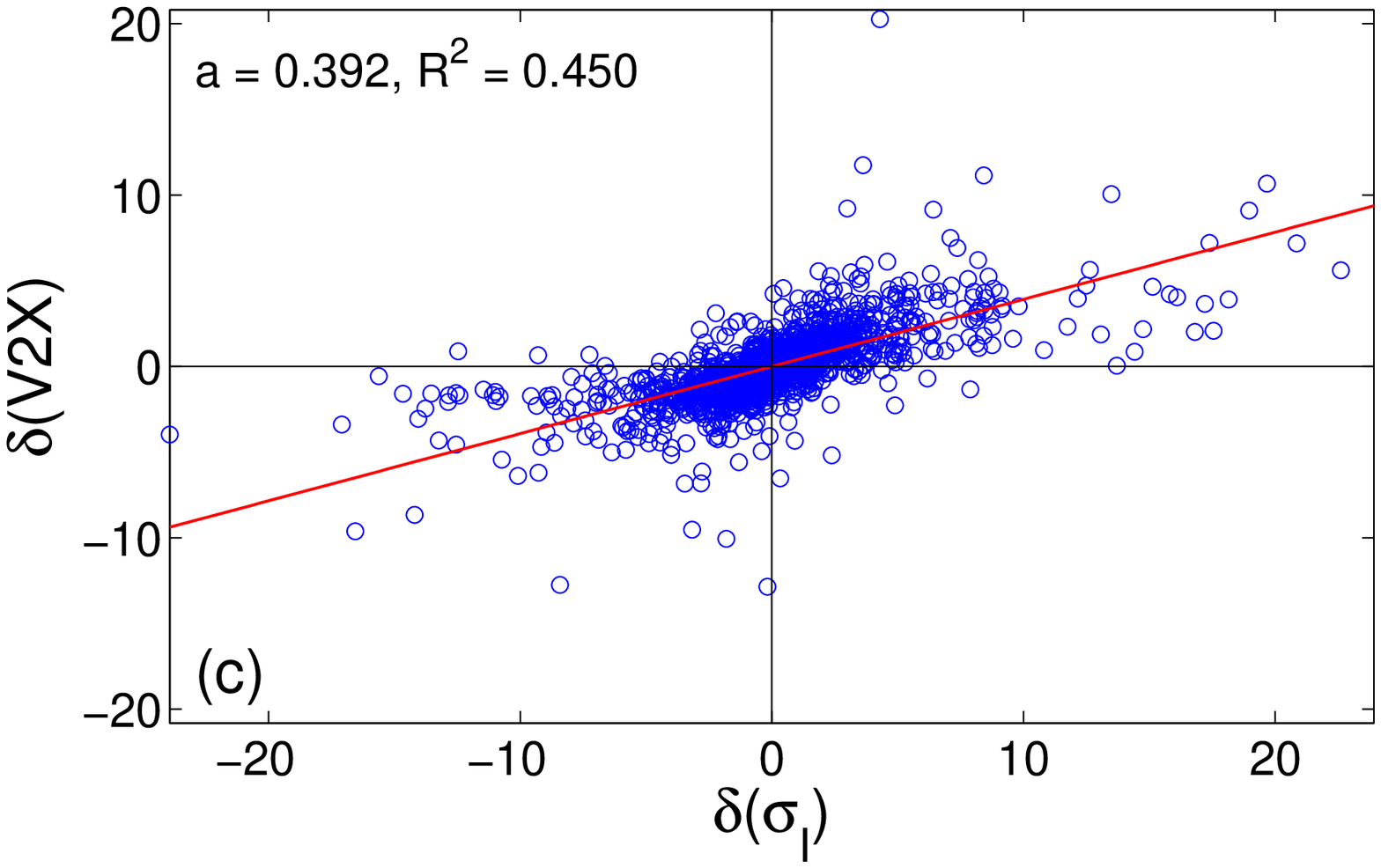}
\includegraphics[width=80mm]{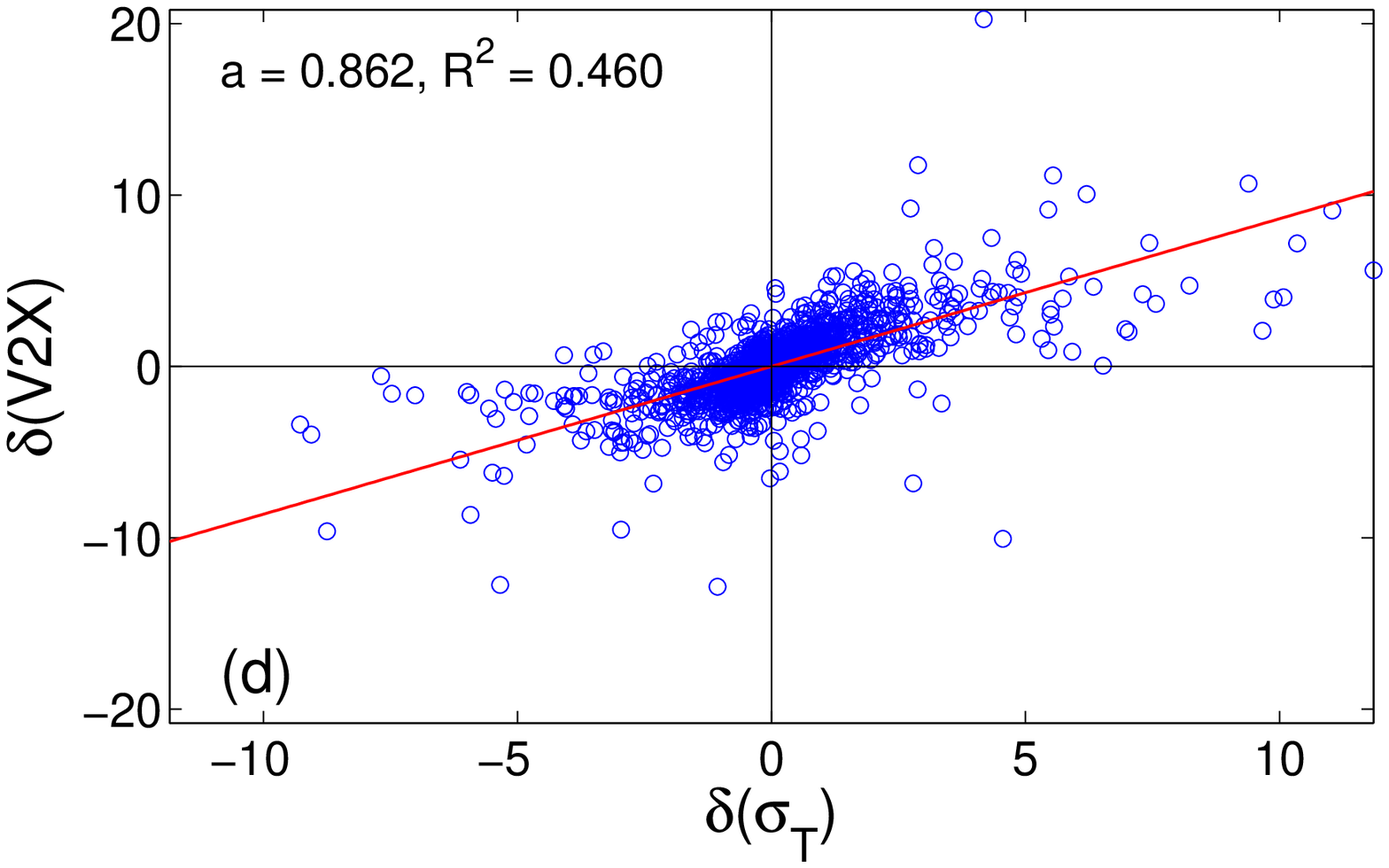}
\end{center}
\caption{
Correlations between increments $\delta(\rm{V2X})$ of the Eurostoxx 50
implied volatility index V2X (vertical axis) and increments of four
different volatility estimators (horizontal axis): (a) a standard
estimator based on the exponential moving average of the squared
returns ($\lambda_\sigma = 1/40$); (b) a standard GARCH estimator
($\omega = 0.0000014$, $\alpha = 0.1064523$, $\beta =
0.8966662$); (c) the reactive volatility estimator without a term
structure; and (d) the reactive volatility estimator with a term
structure.  First, the increments in volatility for both the standard
volatility and GARCH estimators are much more strongly skewed towards
positive values.  Second, the correlations between the V2X and both
reactive volatility estimators are much higher.  Finally, accounting
for the term volatility structure increases the slope of a linear
regression from $0.392$ to $0.862$, with the latter value being close
to $1$.}
\label{fig:correlation}
\end{figure}

Figure \ref{fig:correlation} compares the implied volatility V2X for
Eurostoxx 50 to four different volatility estimators: (i) a standard
estimator with an exponential moving average of squared returns:
\begin{equation}
\sigma_{I,SD}^2(t+1) = (1 - \lambda_\sigma) \sigma_{I,SD}^2(t) + \lambda_\sigma \left(\frac{\Delta I(t+1)}{I(t)}\right)^2
\end{equation}
with the same value of the weighted parameter $\lambda_\sigma = 1/40$;
(ii) a GARCH estimator, $\sigma_{GARCH}(t)$, which is often considered
to be the gold standard; (iii) the reactive volatility estimator
$\sigma_I(t)$ from Eq. (\ref{eq:sigmaIt1}) without a term structure;
and (iv) the reactive volatility estimator $\sigma_T(t)$ from
Eq. (\ref{eq:sigmaT}) with the term structure.  One can see that both
the standard and GARCH estimators have much lower correlations with
V2X than both reactive estimators.  Additionally, the term structure
model for the reactive volatility estimator, as expected, improves the
slope of the linear regression.  Indeed, the slope becomes much closer
to $1$.  The $R^2$ from the linear regression is relatively high
(around $0.45$).  We can therefore consider the model to be nearly as
reactive as the volatility index.

\subsection{Model for combining panic and retarded effects: the case of single stock}

For a single stock, the reactive volatility model relies on an
equation similar to Eq. (\ref{eq:sigmaIt1}) used for the stock index:
\begin{equation}
\label{eq:sigmaIt2}
\sigma_i(t+1) = \tilde{\sigma}_{i}(t+1) \frac{L_i(t+1)}{P_i(t+1)}
\end{equation}
with $L_i(t+1)$ and $\tilde{\sigma}_i(t+1)$ obtained through equations
similar to Eqs. (\ref{eq:L}, \ref{eq:sigmaIt0}), respectively.  The
only difference comes from $\hat{L}_s(t+1)$ in Eq. (\ref{eq:L}), which
applies now to the single stock price $P_i(t)$ instead of the index
price $I(t)$.  Using the Taylor expansion (\ref{eq:L_Taylor}), we
obtain:
\begin{equation}
\sigma_i(t+1) \approx \tilde{\sigma}_i(t+1) \frac{L_{i,s}(t+1)}{P_i(t+1)} \left(1 + \ell \frac{L_f(t+1)-I(t+1)}{L_f(t+1)}\right)
\end{equation}
This formula combines the slow EMA of the single stock price on one
hand, and the fast EMA and the current price of the stock index, on
the other hand.  As a consequence, it adequately extends the leverage
effect from the specific risk case (already accounted in the retarded
volatility model), to the systematic case, which captures the panic
effect.  In what follows, the reactive volatility estimator
$\sigma_i(t)$ from Eq. (\ref{eq:sigmaIt2}) is tested and used to
identify precursors and replicas around extreme events.

Finally although the formula used to determine the level
$L_i(t)$ for the single stock case looks similar to that used to
determine the level $L(t)$ of the stock index, the leverage effect is
different from a practical point of view.  Indeed the leverage effect
for the stock index depends only on the historical stock index price
and is dominated by the fast panic effect; in turn, the leverage
effect for a single stock depends on both the single stock price and
the stock index price, and it is dominated by the slow retarded
effect.  However as soon as the systematic risk becomes dominant as
compared to the specific risk (that happens when the market is highly
stressed), the leverage effect in the single stock starts to be also
dominated by the panic effect.

Our approach manages to combine the panic effect model
introduced in Sect. \ref{sec:stock_index} with the retarded volatility
model by \citet{Bouchaud01}.  This combination allows us to model the
leverage effect for both the systematic and specific risks that occur
in the single stock case.  In that case the level $L(t)$ is that of
the specific risk obtained with the retarded volatility model
modulated by the panic effect in Eq. (\ref{eq:L_Taylor}).  The panic
effect is correlated to the relative excess of the systematic risk
with as compared to fast moving average.

The model manages to adjust its estimation of the volatility in
a instantaneous way at each single stock or index price variation.
Moreover, the adjustment is highly correlated to variations of the
volatility index.  One can expect that the model will manage to
capture, in a reactive way, not only the risk of the stock index but
also the systematic risk of any stock.

\section{Empirical test of the reactive volatility model around extreme events}
\label{sec:empirical}

\subsection{Data analysis}

The database consists of daily price series for the 470 most liquid
European stocks from January 1st 2000 to April 4th 2012.  Let $P_i(t)$
denote the closing price of $i$-th stock at day $t$.  From each price
series, an array of arithmetic returns, $R_i(t) =
\frac{P_i(t) - P_i(t-1)}{P_i(t-1)}$, is constructed.  An extreme event is
said to occur when the absolute arithmetic return $|R_i(t)|$ is three
times larger than the reactive volatility estimator $\sigma_i(t-1)$:
\begin{equation}
\label{eq:cond}
|R_i(t)| > 3 \sigma_i(t-1)
\end{equation}
For each stock, we search for extreme events in the array
$\{R_i(t)\}$.  At each occurrence of an extreme event, we record the
subsequence $\{r(-\Delta), r(-\Delta+1), ... , r(0), r(1), ...,
r(\Delta)\}$ of normalized returns before and after the extreme event
(at day $t$), 
\begin{equation}
\label{eq:rk}
r(k) = \frac{R_i(t+k)}{\sigma_i(t+k-1)}  \qquad  (k = -\Delta ... \Delta) 
\end{equation}
where $\Delta$ is a fixed subsequence length with $\Delta = 9$ trading
days.\footnote{The length is fixed to 9 trading days after having
considered up to $\pm 100$ days; empirically, after 9 days, the
marginal gain in precision can be neglected.}  The subsequence
$\{r(k)\}$ characterizes the behavior of a stock before and after an
extreme event, which is identified by the magnitude of $r(0)$ (we drop
the index $i$ because the extreme events will be analyzed for all
stocks together).  Repeating this procedure for each stock generates a
database of 10,213 extreme events.  Note that if two (or more) extreme
events occurred within $\Delta$ days, only the first one is retained,
while any later events are ignored.  It is worth emphasizing that the
above definition of an extreme event is purely conventional, as is the
choice for the threshold $3\sigma_i(t-1)$.  Note also that if stock
returns were Gaussian, the number of extreme events would be much
smaller than what we observe because the probability of a Gaussian
return larger than $3\sigma$ is $0.0027$.

A closer look into the statistical properties of stock returns around
extreme events requires distinguishing the systematic risk from the
specific risk.  The systematic risk mainly affects the stock indices
(or even every stock during stress conditions), while the specific
risk mainly affect single stocks.  More precisely, the database
records are split into four groups that are denoted ``systematic
positive'' (SyP), ``systematic negative'' (SyN), ``specific positive''
(SpP) and ``specific negative'' (SpN).  The division into positive and
negative groups is determined by the sign of the extreme normalized
return $r(0)$.  The division into systematic and specific groups is
decided by the condition on the Eurostoxx index $I(t)$ at the day $t$
of an extreme return: if $|\Delta I(t)|$ exceeds $3\%$, the extreme
event is categorized as systematic, otherwise as specific.  Both
systematic groups (SyP \& SyN) contain the records of extreme returns
from individual stocks that are affected by large stock index returns,
representing extreme events for the entire market.  The specific
groups (SpS \& SpN) contain the records of extreme returns that are
specific to individual stocks.  The database of extreme events
contains 903 systematic positive records, 1,046 systematic negative
records, 5,135 specific positive records and 3,129 specific negative
records.

To qualitatively separate the possible sources of deviations (accuracy
of the reactive volatility estimator and return correlations), a
second database of ``non-extreme'' events is constructed with the same
structure, in which dates $t$ are chosen randomly (without the
selective condition (\ref{eq:cond})).  In this case, no strong
correlations between successive returns are expected, and the reactive
volatility estimator is expected to accurately capture the
fluctuations of the stock price.

\subsection{ Empirical results }

The behavior of the reactive volatility model around extreme events
is characterized by the following function:
\begin{equation}
\label{eq:qk}
q_k = \sqrt{\langle r^2(k) \rangle} - 1     \qquad (k = -\Delta \ldots \Delta )
\end{equation}
where the arithmetic average $\langle ... \rangle$ is taken over all
records in the chosen group (SyP, SyN, SpP, SpN).  For the idealized
case in which the returns are uncorrelated and the reactive volatility
estimator $\sigma_i(t)$ is exact, the average $\langle r^2(k) \rangle$
of normalized returns should be equal to $1$, so that $q_k$ would be
$0$, except for $k = 0$.  In other words, the smallness of deviations
of $q_k$ from $0$ characterizes the accuracy of the volatility
estimator.

\begin{table}
\begin{center}
\begin{tabular}{| c | c | c || c | c |}  \hline
\multirow{2}{*}{Group} & \multicolumn{2}{|c||}{Before} & \multicolumn{2}{c|}{After} \\  \cline{2-5}
                       & Standard & Reactive & Standard & Reactive \\  \hline
Systematic positive & 0.76 & 0.34 & 0.52 & 0.35 \\
Systematic negative & 0.55 & 0.21 & 1.02 & 0.54 \\
Specific positive & 0.54 & 0.11 & 0.90 & 0.31 \\
Specific negative & 0.54 & 0.22 & 1.13 & 0.45 \\  \hline
\end{tabular}
\end{center}
\caption{
The level of deviations $q_k$ is averaged over 9 days before and after an
extreme event. For all groups of extreme events, this level is
significantly higher for a standard volatility estimator than for the
reactive volatility model because the latter captures the panic effect
for the systematic groups and the retarded effect for the specific groups. 
Note also that the recovery after an extreme event
takes, on average, $2.95$ days for the reactive volatility model, as
opposed to $5.29$ days for a standard estimator (both times are
estimated from an exponential fit of $q_k$ after an extreme event; see
Fig. \ref{fig:returns}). As a consequence, the reactive volatility
model is qualified as more robust in response to extreme events. }
\label{tab:levels}
\end{table}

\begin{figure}%
\begin{center}
\includegraphics[width=80mm]{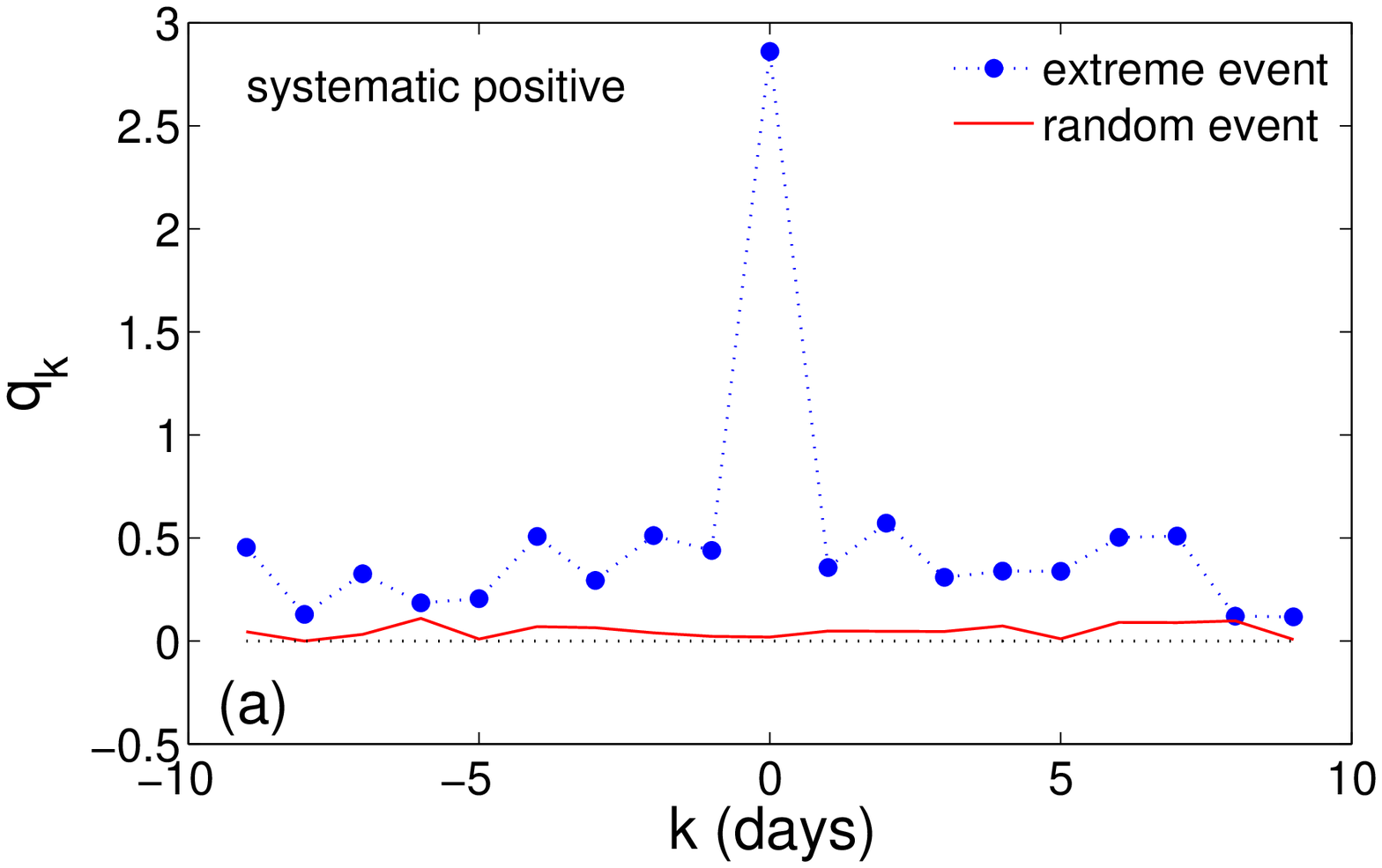}
\includegraphics[width=80mm]{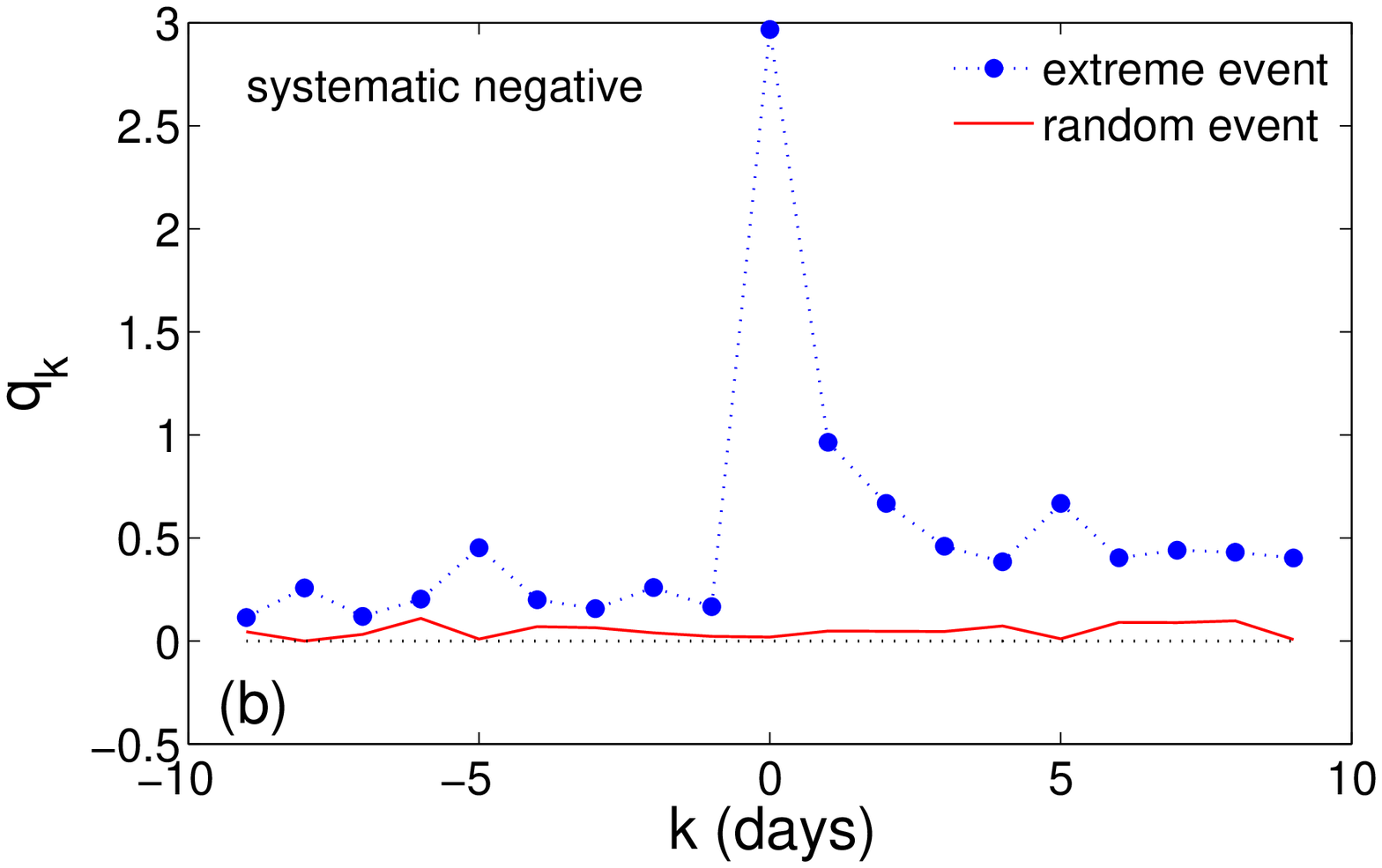}
\includegraphics[width=80mm]{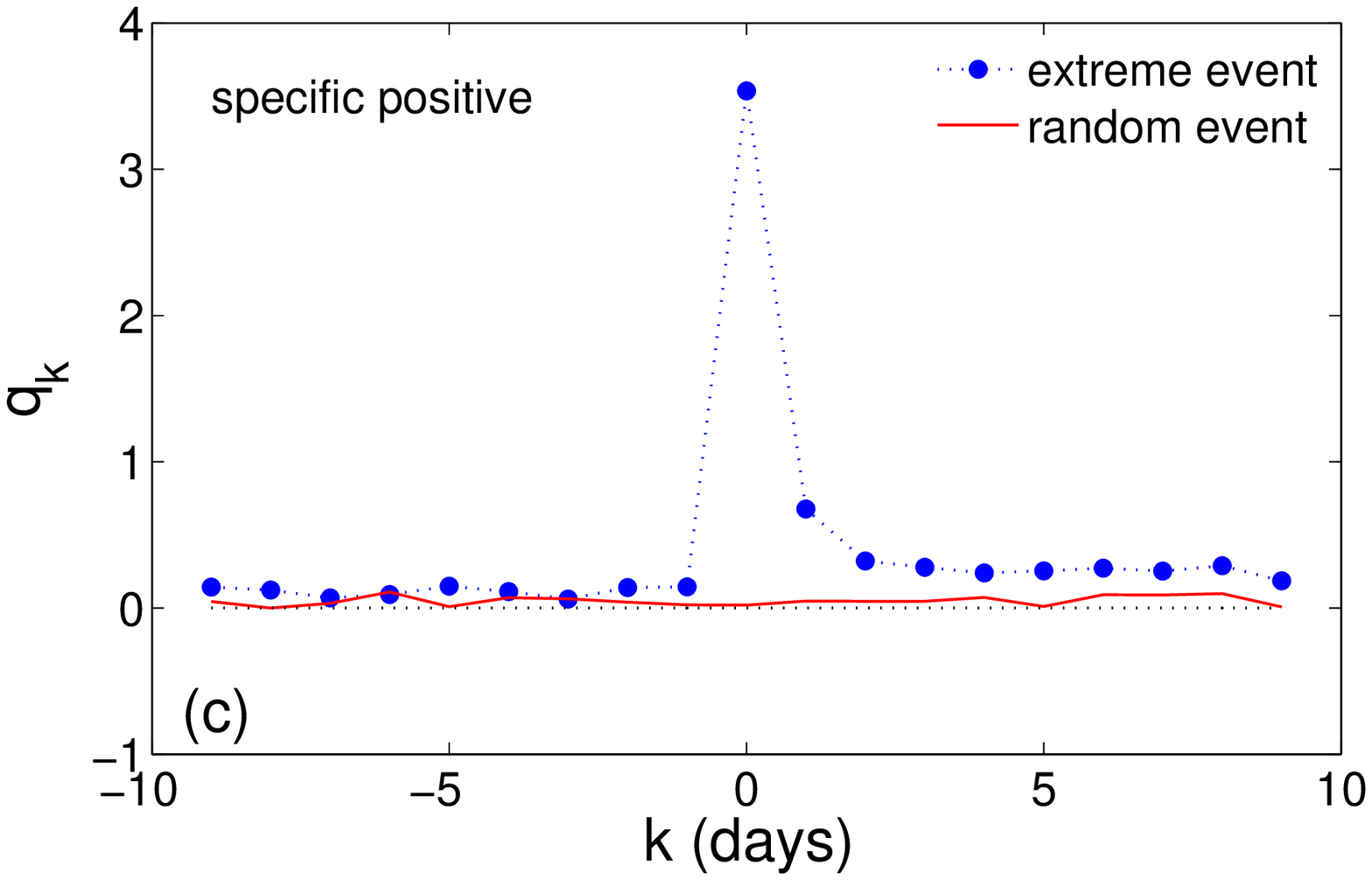}
\includegraphics[width=80mm]{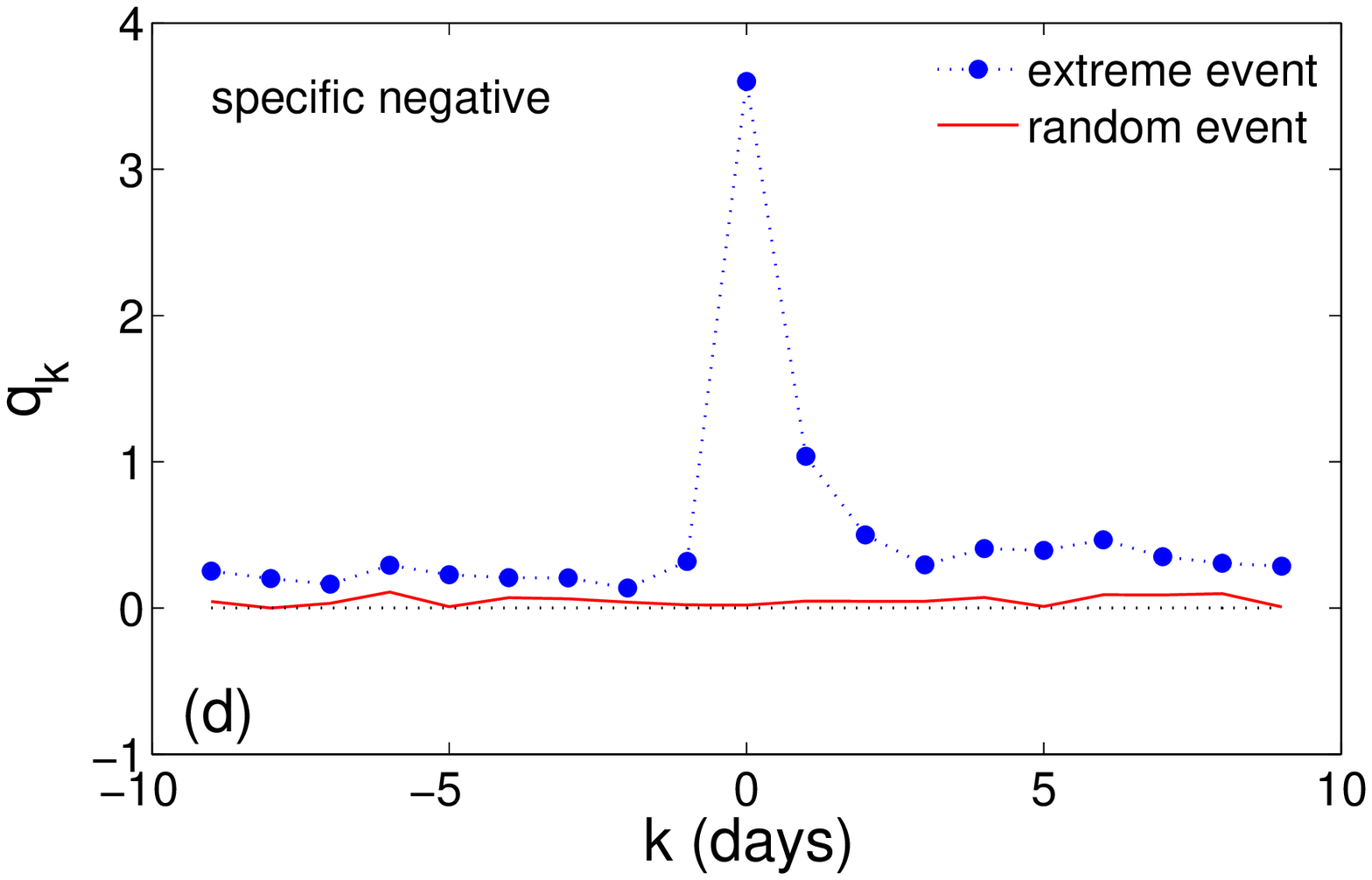}
\end{center}
\caption{
Distribution of $q_k$ around an extreme event for the reactive
volatility model.  Extreme returns are split into four groups: (a)
SyP, (b) SyN, (c) SpP and (d) SpP.  A stock return is termed extreme
when it exceeds threefold the empirical volatility $\sigma_i(t-1)$
from the reactive volatility model.  The selected interval is $\pm 9$
days.}
\label{fig:returns}
\end{figure}

\begin{figure}%
\begin{center}
\includegraphics[width=80mm]{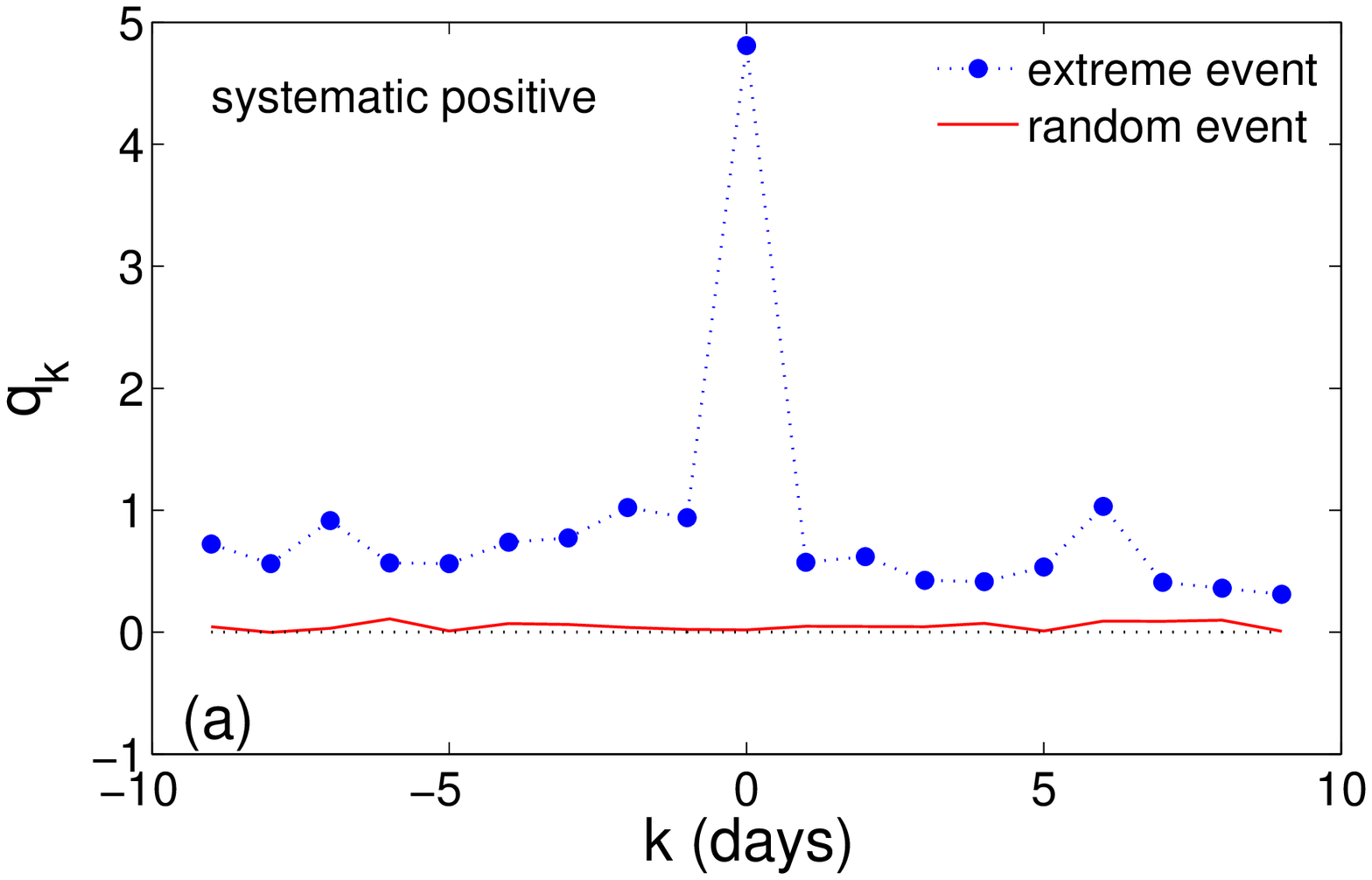}
\includegraphics[width=80mm]{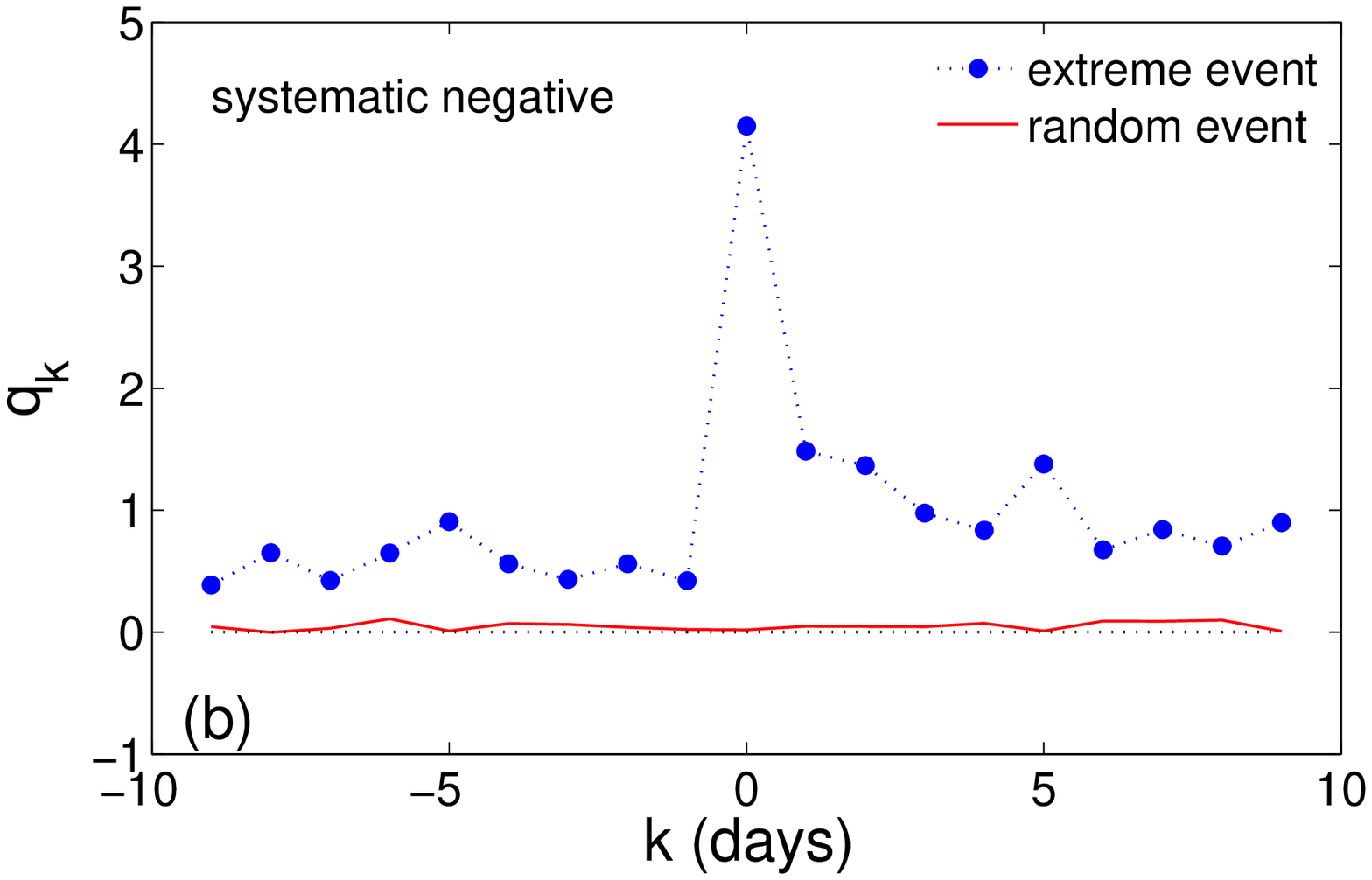}
\includegraphics[width=80mm]{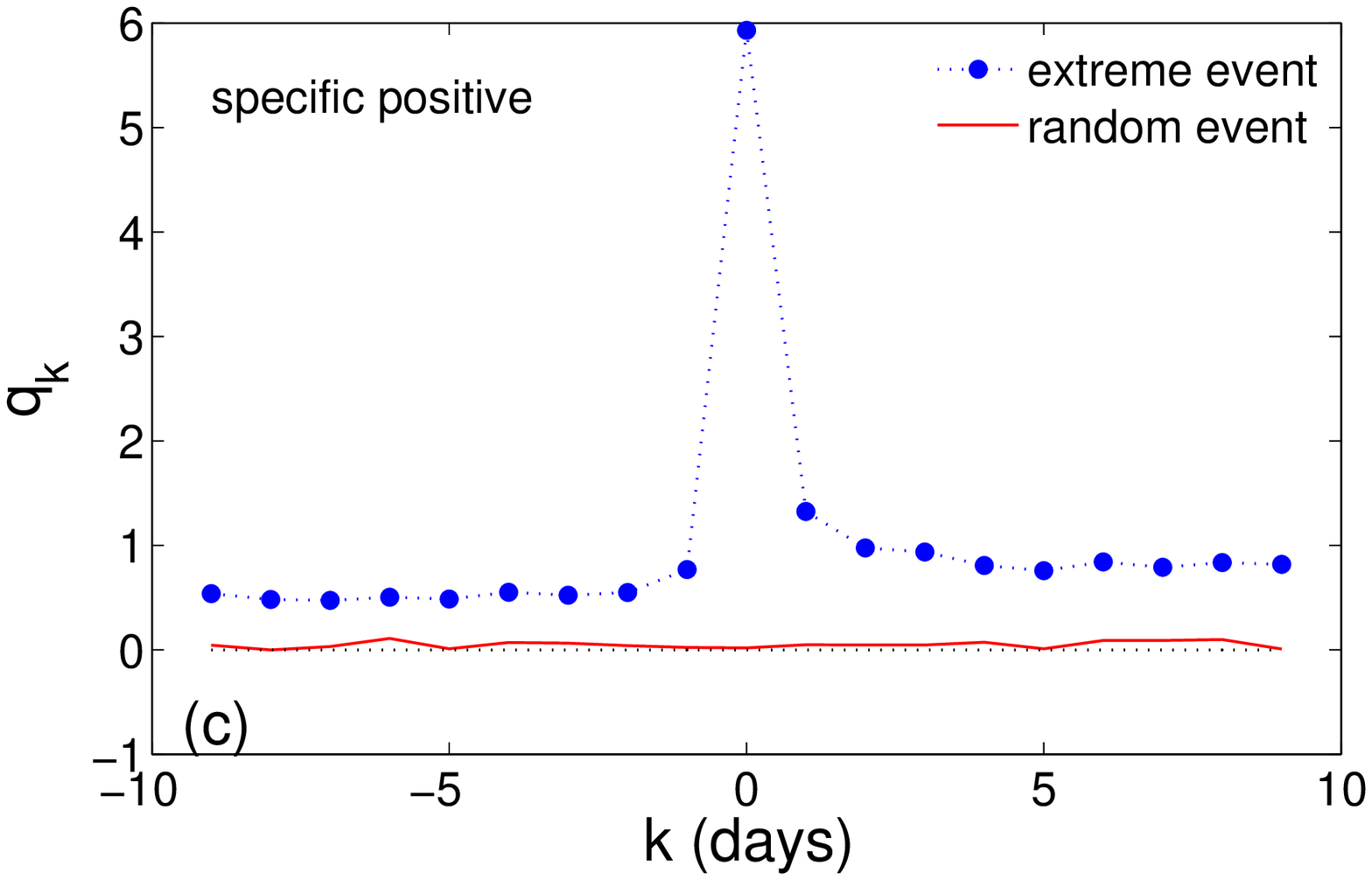}
\includegraphics[width=80mm]{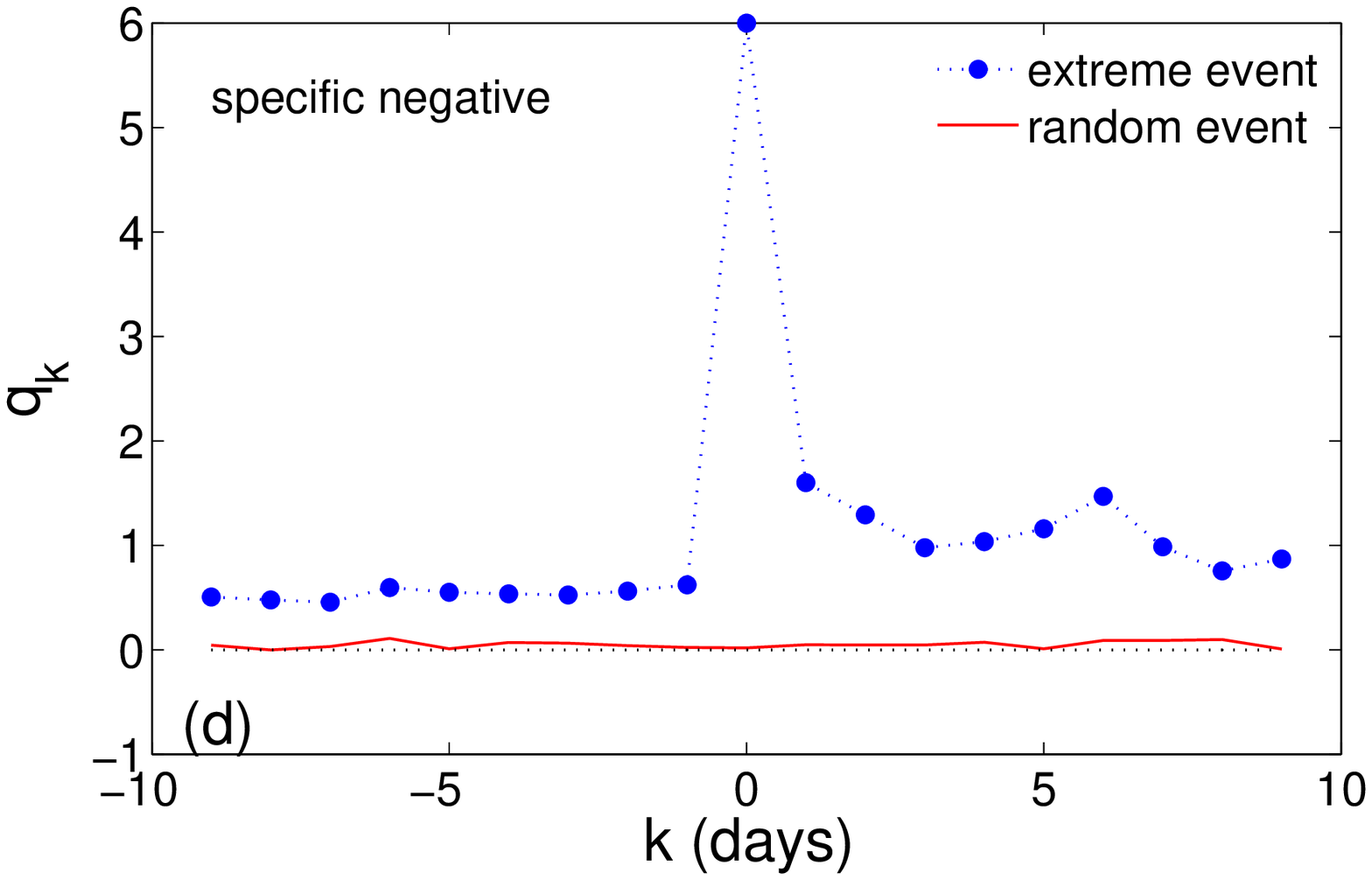}
\end{center}
\caption{
Distribution of $q_k$ around an extreme event for a standard
volatility estimator.  Extreme returns are split into four groups: (a)
SyP, (b) SyN, (c) SpP and (d) SpN.  A stock return is termed extreme
when it exceeds threefold the empirical volatility
$\sigma_{i,SD}(t-1)$ from a standard volatility estimator.  The
selected interval is $\pm 9$ days.}
\label{fig:returns_SD}
\end{figure}

Table \ref{tab:levels} summarizes the average characteristics obtained
for both estimators.  The level of deviations $q_k$ is averaged over 9
days before and after an extreme event, as illustrated in
Figs. \ref{fig:returns}, \ref{fig:returns_SD}.  For all groups of
extreme events, this level is significantly higher for the standard
volatility estimator than for the reactive volatility model because
the latter captures the panic effect for the systematic groups and the
retarded effect for the specific groups.  One can conclude that the
reactive volatility model is more robust near extreme events, and its
recovery after a shock is faster.

Figure \ref{fig:returns} shows $q_k$ for the four groups.  Full
circles represent $q_k$ from the database of extreme events, while the
solid line is used as a reference level from the database of random
events (extreme or not).  The solid line is close to $0$, which
indicates that the reactive volatility model is accurate for ordinary
days (without extreme events).  For comparison,
Fig. \ref{fig:returns_SD} shows the quantities $q_k$ computed in
Eq. (\ref{eq:qk}) by replacing $\sigma_i(t)$ in Eq. (\ref{eq:rk}) with
an empirical measure of volatility, $\sigma_{i,SD}(t)$, which is
obtained from a standard volatility estimator with an exponential
moving average:
\begin{equation}
\sigma_{i,SD}^2(t+1) = (1-\lambda_\sigma) \sigma_{i,SD}^2(t) + \lambda_\sigma [R_i(t+1)]^2
\end{equation}
with the same value of the weighting parameter $\lambda_\sigma =
1/40$.

Let us now take a closer look at the results of
Fig. \ref{fig:returns}.  For systematic groups, there are significant
deviations of $q_k$ from $0$ before and after an extreme event.  In
other words, the normalized returns before and after an extreme event
are significantly larger than typical normalized returns.  Prior to an
extreme event, the normalized returns increase very slowly and become
significantly larger than typical normalized return 30 days before the
event.  The relaxation time after the extreme event is on contrary
much faster.

In the systematic positive group, the observation of relatively strong
precursors might be partly explained by political or monetary
decisions made in reaction to earlier market instabilities.

For the systematic negative group, the observation of excitements
before an extreme event is less intuitive.  These variations could
come from the possibility that some investors anticipate the release
of very bad economic news earlier than others.  After an extreme
event, large renormalized returns are naturally expected as the market
relaxes after this event. It means that investors should be worried
about replicas after the initial extreme event and that most models
consistently underestimate risk.

Figures \ref{fig:returns}c, \ref{fig:returns}d for both specific
groups indicate less significant deviations of $q_k$ from $0$ before
an extreme event.

In the specific positive group, the observation of an extreme positive
return can be caused by announcements of corporate decisions (mergers
or acquisitions).  Because these corporate decisions remain strictly
confidential and are difficult to anticipate, one expects to observe
very weak precursors and small values of $q_k$ for negative $k$.
After the day of announcement, the stock returns may experience a
rapid relaxation (one or two days) towards the normal level of small
$q_k$ for positive $k$.

For the specific negative group, the observation of an extreme event
can be caused by an announcement of corporate decisions related to bad
economic results for the company (profit warning, bankruptcy, or
downgrade).  This kind of news is partly anticipated by the market
through rumors.  One therefore can identify stronger precursors of
extreme events.  Because the situation of the company remains tenuous,
stronger replicas are also identified.

The mechanisms of these precursors in every case seem plausible
especially since the volatility forecasts are not particularly low
when these relatively large events may happen.  Indeed we compare the
whole sample and the conditional sample defined by the days prior to
any extreme event until 9 days. The average of the volatilities
forecasted by our volatility model is $31\%$ in the conditional sample
and remains close to that estimated for the whole sample ($33\%$).  We
also find similar results for the standard deviations of returns.  We
manage to identify {\it statistically} the presence of precursors of
extreme events in each case, but the precursors remain too weak to be
detected one by one to forecast the occurrence of an extreme event.
Indeed the predictive power is low and the model cannot be used to
forecast any crash or extreme event. For example, if a warning signal
emerges each time the realized volatility exceeds twice the level
predicted by the reactive volatility model for a period of 9 days,
$96.68\%$ of the extreme events will be missed, while $56.8\%$ will be
false warning signals.

\begin{figure}
\begin{center}
\includegraphics[width=80mm]{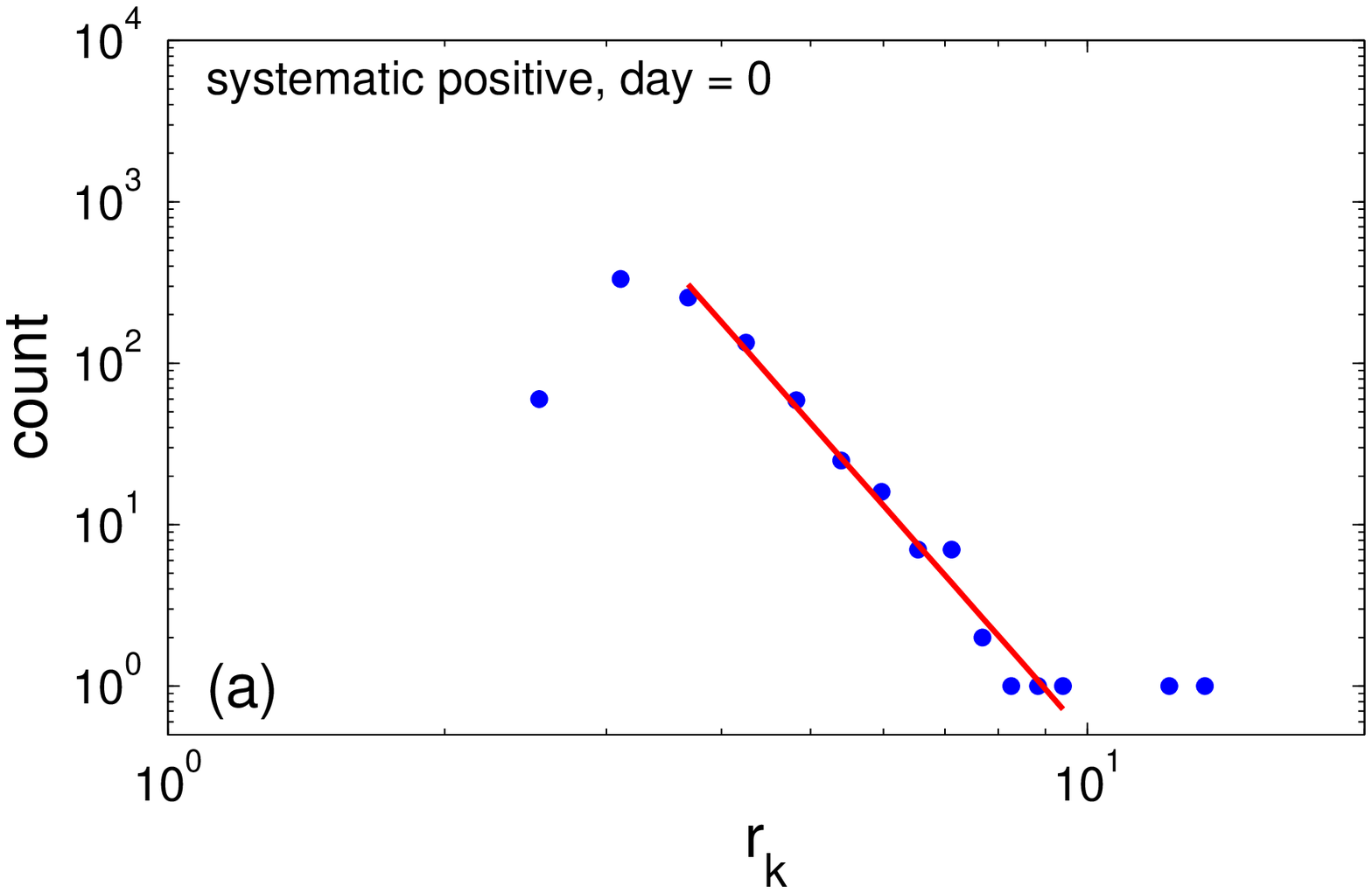}
\includegraphics[width=80mm]{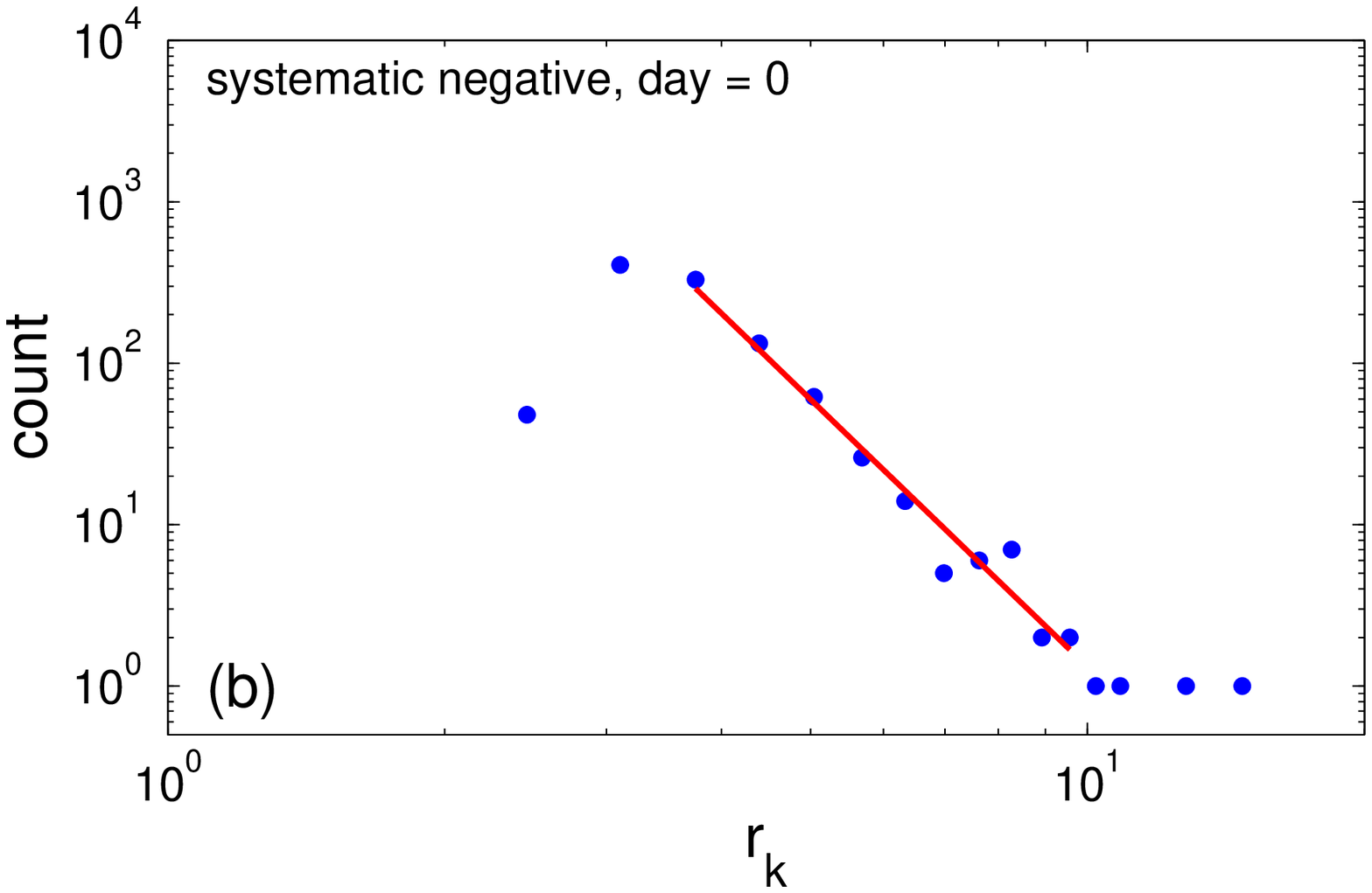}
\includegraphics[width=80mm]{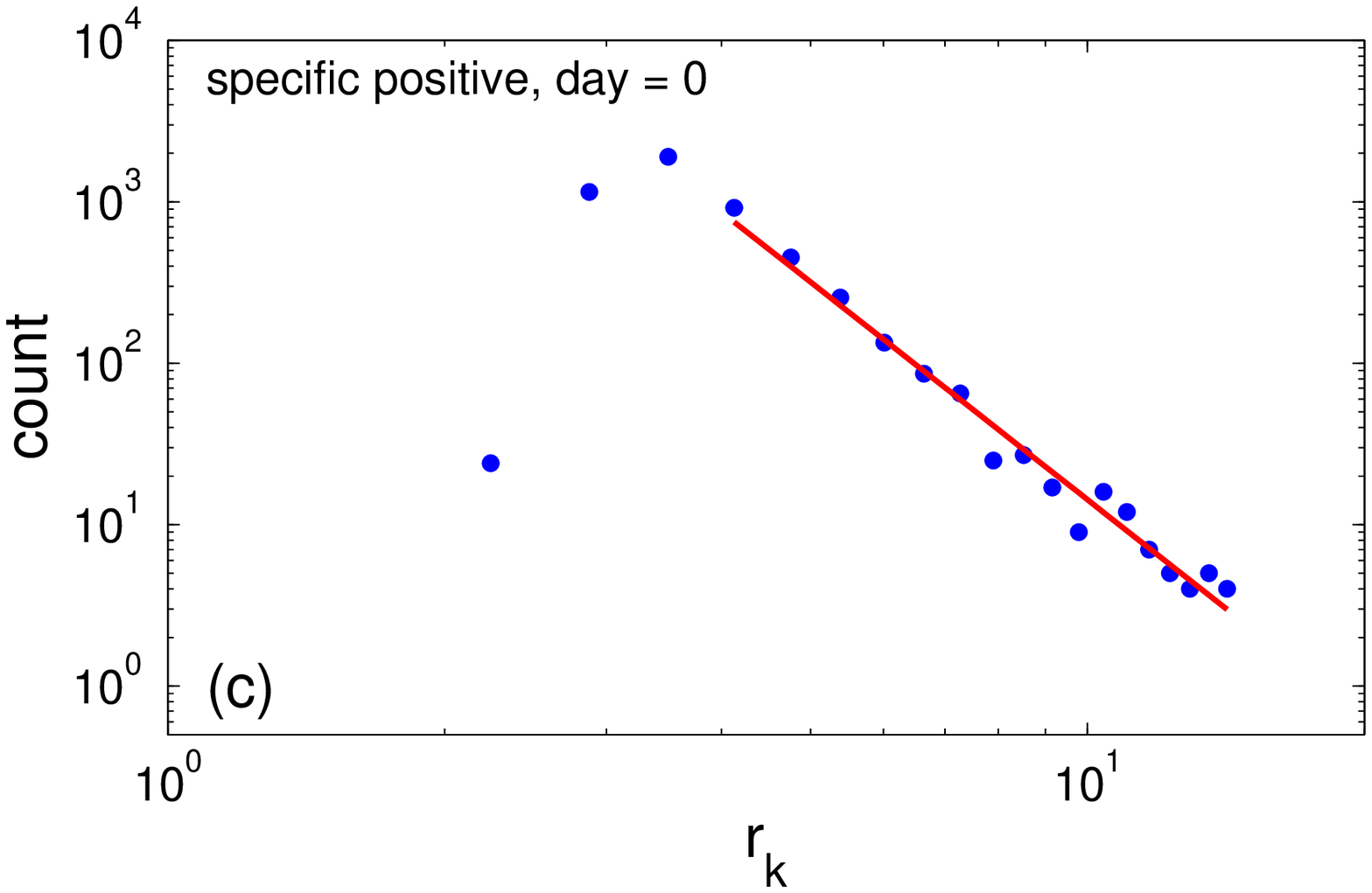}
\includegraphics[width=80mm]{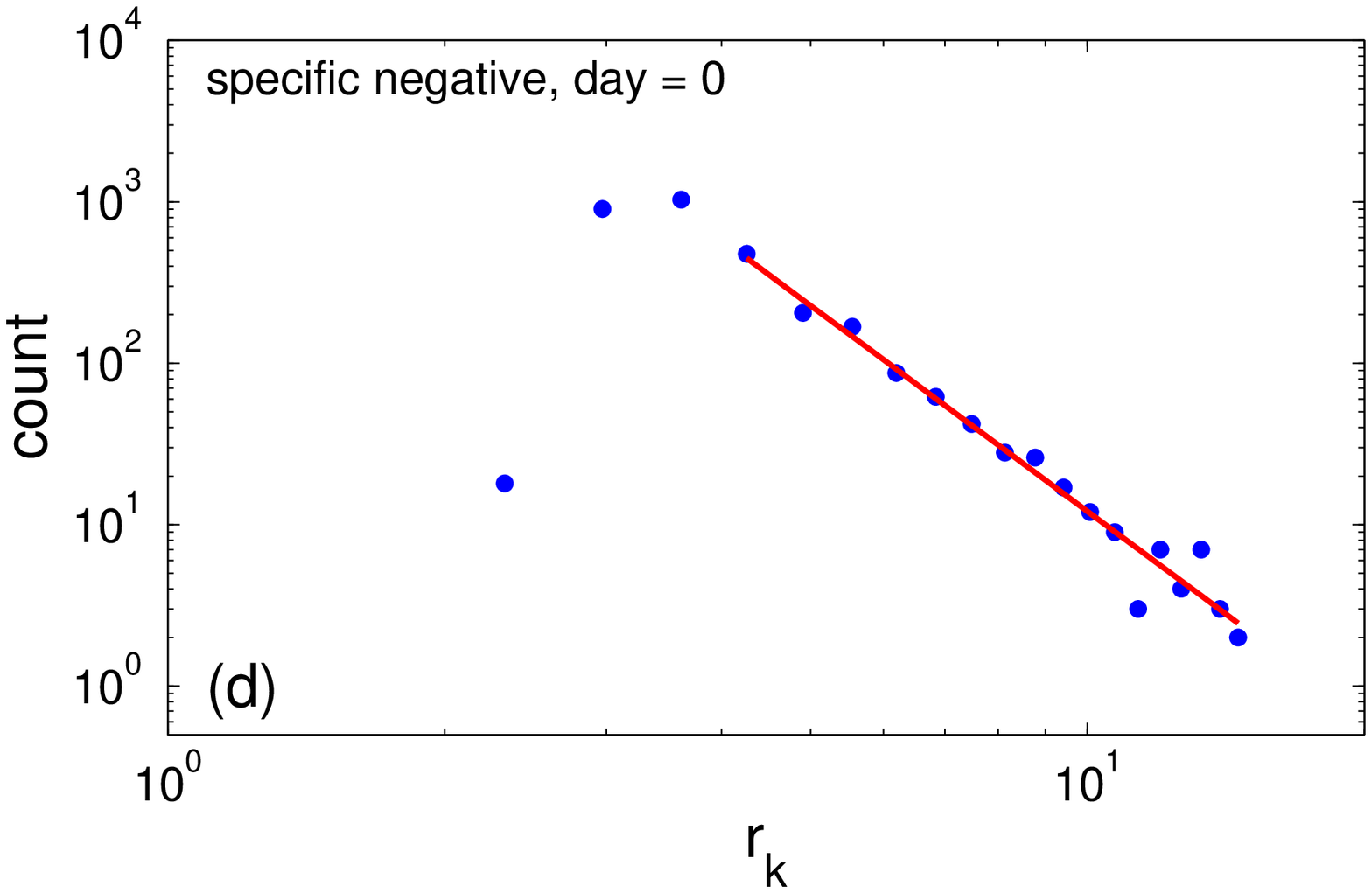}
\end{center}
\caption{
Asymptotic behavior of the probability density $p(z)$ of extreme
returns normalized with the reactive volatility model} for the
four groups: (a) SyP, (b) SyN, (c) SpP and (d) SpN.  In all cases,
power law decays $p(z)\propto z^{-1-\alpha}$ are observed, with the
exponents $5.5$, $4.5$, $3.5$ and $3.2$ for the systematic positive,
systematic negative, specific positive and specific negative groups,
respectively.  For the systematic groups, the exponent is greater than
$4$, indicating the existence of kurtosis, while for the specific
groups, the exponent is less than $4$.
\label{fig:exponent}
\end{figure}

Similar plots in Fig. \ref{fig:returns_SD} for a standard volatility
estimator display a less reactive measure around extreme events.

Figure \ref{fig:exponent} shows the asymptotic behavior of the
probability density $p(z)$ of extreme normalized returns for the four
groups.  In all cases, power law decays $z^{-1-\alpha}$ are observed,
with the exponent $\alpha$ equal to 5.5, 4.5, 3.5 and 3.2 for the SyP,
SyN, SpP and SpN groups, respectively.  For the systematic groups, the
exponent is greater than $4$, indicating the existence of kurtosis.
The renormalization of returns with the reactive volatility model,
instead of using the standard EMA volatility estimate, has managed
to increase the exponent from $3$ to $5$, which means that this model
is able to capture most of the extreme events.  For the specific
groups, the exponent remains around $3$.  Most of these extreme risks
come from very specific news (for example, a takeover offer), and most
of the time, when the price jumps, the volatility does not change.

\section{Conclusion}
\label{sec:conclusion}

We developed a new volatility model, easy to implement, that
includes a leverage effect whose return-volatility correlation
function fits to empirical observations.  In addition, the model is
able to capture both the panic effect induced by the systematic risk
and the retarded effect induced by the specific risk.  The model is
shown to be as reactive as the implied volatility, which is an
improvement over other models.  To test the robustness of the reactive
volatility model near extreme events, an empirical study is performed
on 470 the most liquid European stocks from January 1st 2000 to April
4th 2012.  The reactive volatility model is used to renormalize daily
returns, among which extreme events are identified and split into four
groups: systematic positive, systematic negative, specific positive
and specific negative.  Our results suggest that the market shocks are
better assimilated into the reactive volatility model.  Moreover, the
model identifies statistically the presence of precursors and
replicas.  The model captures much of the extreme systematic risk and
a significant part of the extreme specific risk.  Future research will
include an application of the reactive volatility model to estimate
market beta, the aggregation of risk and VaR of a Long/Short
portfolio.

\pagebreak



\end{document}